\documentclass[10pt,twocolumn,english,preprint,prl,amsfonts,amssymb,byrevtex,tightenlines,nobalancelastpage]{revtex4}
\usepackage[T1]{fontenc}
\usepackage[latin9]{inputenc}
\usepackage{amsmath}
\usepackage{graphicx}

\makeatletter
\@ifundefined{textcolor}{}
{%
 \definecolor{BLACK}{gray}{0}
 \definecolor{WHITE}{gray}{1}
 \definecolor{RED}{rgb}{1,0,0}
 \definecolor{GREEN}{rgb}{0,1,0}
 \definecolor{BLUE}{rgb}{0,0,1}
 \definecolor{CYAN}{cmyk}{1,0,0,0}
 \definecolor{MAGENTA}{cmyk}{0,1,0,0}
 \definecolor{YELLOW}{cmyk}{0,0,1,0}
 }

\providecommand{\U}[1]{\protect\rule{.1in}{.1in}}

\makeatother

\usepackage{babel}

\begin{document}

\title{Magnetic Field Tuned Quantum Phase Transition in the Insulating Regime
of Ultrathin Amorphous Bi Films}

\author{Yen-Hsiang Lin and A. M. Goldman}

\affiliation{School of Physics and Astronomy, University of Minnesota, 116 Church
St. SE, Minneapolis, MN 55455, USA}
\begin{abstract}
A surprisingly strong variation of resistance with perpendicular magnetic
field, and a peak in the resistance vs. field, $R(B)$ has been found
in insulating films of a sequence of homogeneous, quench-condensed
films of amorphous Bi undergoing a thickness-tuned superconductor-insulator
transition. Isotherms of magnetoresistance, rather than resistance,
vs. field were found to cross at a well-defined magnetic field higher
than the field corresponding to the peak in $R(B)$. For all values
of B,{\normalsize{} $R(T)$ was found to obey an Arrhenius form. At
the crossover magnetic field the prefactor }became equal to the quantum
resistance of electron pairs, $h/4e^{2}$, and the activation energy
returned to its zero field value. These observations suggest that
the crossover is the signature of a quantum phase transition between
two distinct insulating ground states, tuned by magnetic field. 
\end{abstract}
\maketitle
Superconductor-insulator (SI) transitions of disordered two-dimensional
(2D) conductors have been studied extensively for about two decades
because they offer the opportunity to investigate a wide variety of
quantum phenomena \cite{Markovic}. Of particular interest are transitions
of strongly disordered films, tuned by a perpendicular magnetic field.
The dirty-boson picture was proposed to describe the magnetic field
tuned transition from superconductivity. In this picture the insulator
consists of Bose-condensed, field-induced vortices and localized Cooper
pairs \cite{Fisher}. An early experiment by Paalanen, Hebard and
Ruel reported a \textit{peak} in the magnetoresistance of InO$_{x}$
films on the insulating side of the SI transition \cite{Hebard}.
The behavior of the Hall resistance at fields close to the peak field
led these authors to suggest that there was a crossover from the state
proposed by Fisher in which there are localized Cooper pairs, to one
in which transport is dominated by single-particle excitations. They
referred to this as crossover between Bose and Fermi insulators \cite{Hebard}.
This peak in $R(B)$ in the insulating regime of the field-tuned SI
transition has been the subject of numerous investigations in recent
years. With improvements in sample fabrication procedures and the
introduction of new materials, changes of resistance of several orders
of magnitude have been reported \cite{Gantmakher1,Lee1,Sambandamurthy,Baturina,Steiner}.
Also of interest in the present context are observations of Arrhenius
activated behavior, \textit{i.e}., a hard gap in InO$_{x}$ and TiN$_{x}$
films both in zero field in insulating films, and in magnetic fields
on the insulating side of the SI transition \cite{Shahar,Kowal,Baturina}.
In this letter we report an apparent perpendicular magnetic field
tuned quantum phase transition between two separate insulating ground
states in thin films on the insulating side of the disorder or thickness-tuned
superconductor-insulator transition. A central piece of evidence for
this assertion is that isotherms of magnetoresistance ($MR$) defined
as $\left[R(B,T)-R(0,T)\right]/R(0,T)$ cross at a well-defined magnetic
field higher than that corresponding to the peak in $R(B)$. 

The data employed in the present work were obtained from studies of
homogeneous amorphous Bi (\textit{a-}Bi) films that were grown by
quench-condensation \textit{in situ} at liquid helium temperatures
on (100) SrTiO$_{3}$(STO) single-crystal substrates precoated \textit{in
situ} with a 15Å underlayer of amorphous Sb (\textit{a-}Sb). Films
grown by deposition onto substrates held at liquid helium temperatures
and pre-coated \textit{in situ} with thin underlayers of%
\begin{figure}
\includegraphics{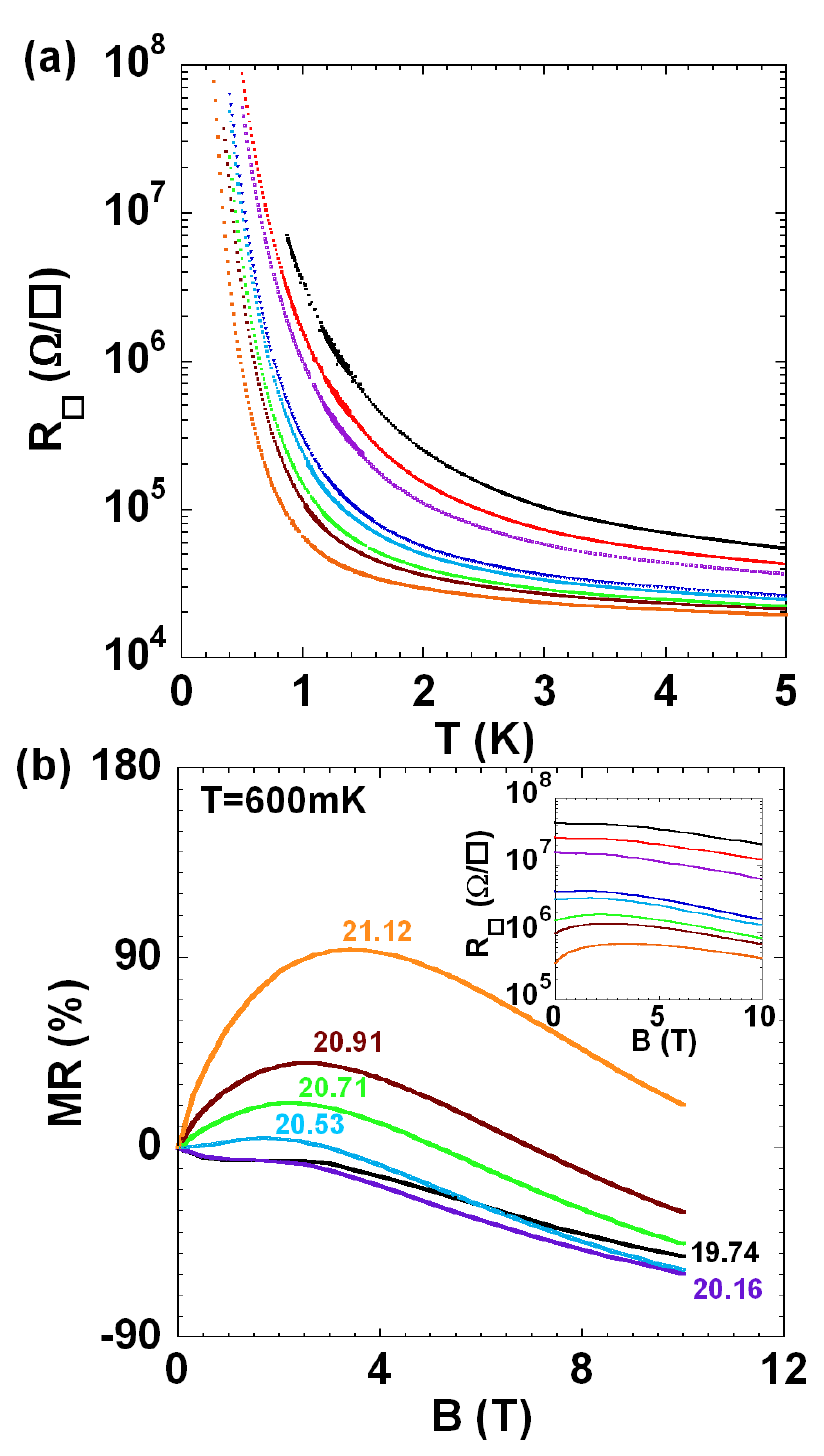}

\caption{(a) Zero field resistance vs. temperature of a sequence of nominally
homogeneous \textit{a-}Bi films with thicknesses from 19.74Å (top)
to 21.12Å (bottom) in average nominal increments of 0.2Å. Notice that
the resistances of these films monotonically increase with decreasing
temperature and do not exhibit the local minima found in nominal granular
films. (b) MR as a function of field at 600mK in films of different
thicknesses. The labels are thicknesses in units of Angstroms. The
inset is the original sheet resistance vs. magnetic field at 600mK,
again for films of different thicknesses. }
 
\end{figure}
either \textit{a-}Ge or \textit{a-}Sb are known to be homogeneous
\cite{Strongin}. The underlayers have zero conductance within instrumental
resolution. The experiments involve repeated cycles of deposition
and measurement carried out in a dilution refrigerator system designed
to study the evolution of electronic properties with film thickness
\cite{Hernandez}. All the measurements were carried out using a four-terminal
configuration employing a DC current source with currents in the linear
regime of the current-voltage (I-V) characteristic. 

Representative examples of the evolution of $R(T)$ with thickness
of several insulating films of \textit{a}-Bi films are shown in Fig.
1(a). Representative data of $R(B)$ and the field dependence of the
$MR$ at 600mK in films ranging in thickness from 19.74Å to 21.12Å
are presented in Fig. 1(b). Peaks in R(B) are observed in films thicker
than 20.53Å. The values of the magnitudes of the peaks in $R(B)$
and the fields at the peaks both increase with film thickness. It
is important to note that large peaks in $R(B)$, at fields above
the critical field of the SI transition have not been previously reported
for superconducting films grown on substrates with \textit{a-}Ge or
\textit{a-}Sb underlayers. On the other hand in the case of nominally
granular quench-condensed films, grown on substrates that are not
precoated, $R(B)$ increases dramatically with increasing field, rising
to values several orders of magnitude higher than the normal resistance\cite{YenHsiangLin}.
Such films are also exhibit nonmonotonic variations of $R(T)$ which
are not found in precoated films. Giant magnetoresistance peaks have
been found in studies of films quench-condensed onto substrates perforated
with nanometer scale arrays of holes \cite{Nguyen}.

\begin{figure}
\includegraphics{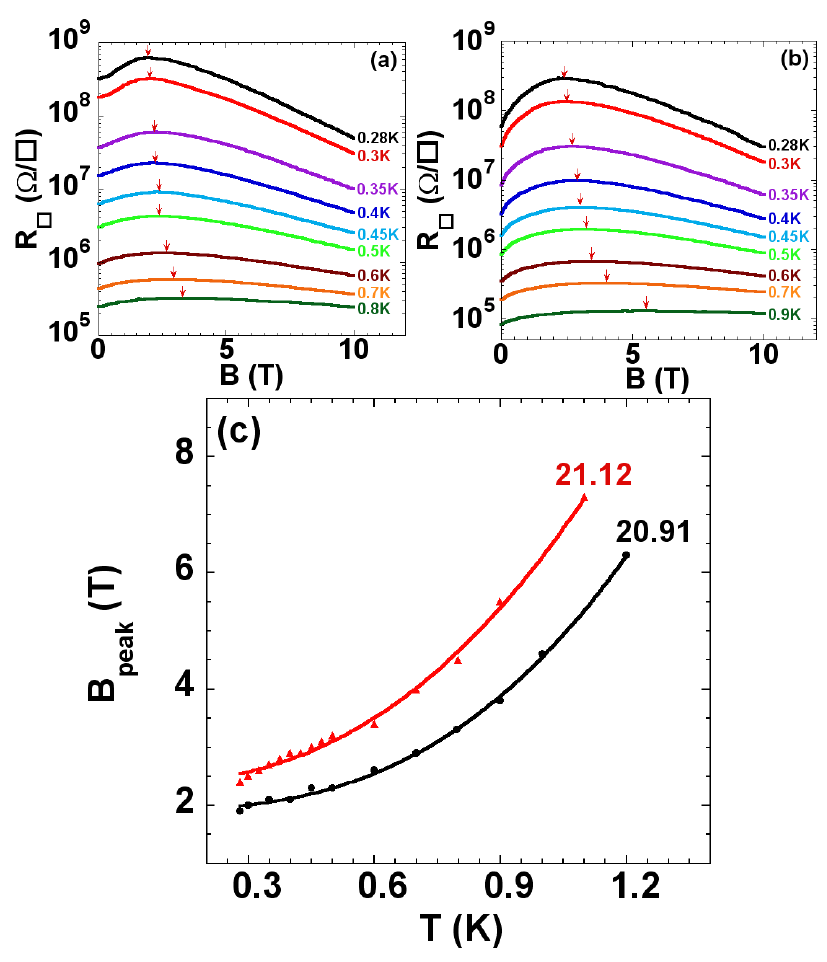}

\caption{Sheet resistance vs. perpendicular magnetic field at different temperatures
for the (a) 20.91Å and (b) 21.12Å thick films. The magnetic field
at the MR peak vs. temperature is plotted in (c). The two thicknesses
are labeled with numbers whose units are Angstroms. The arrows in
(a) and (b) indicate the resistance peaks.}

\end{figure}

We now turn to the temperature dependence of $R(B)$ and the $MR$
for films of specific thicknesses. Representative data of $R(B,T)$
for films, 20.91Å and 21.12Å thick, are presented in Figs. 2(a) and
2(b). The peak height becomes higher with decreasing temperature,
which is consistent with results reported for InO$_{x}$ and TiN$_{x}$
films. The peak field, $B_{peak}$, is a function of temperature and
can be fit with the form, $B_{peak}=B_{0}+\alpha T^{\beta}$ over
the range of temperatures studied, as shown in Fig. 2(c). From the
measurements, $B_{0}=1.92\pm0.04$, $\alpha=2.63\pm0.06,$ and $\beta=2.78\pm0.10$
for the 20.91Å thick film, and $B_{0}=2.37\pm0.07$, $\alpha=3.791\pm0.09,$
and $\beta=2.41\pm0.13$ for the 21.12Å thick film. It is unclear
as to whether any of the theoretical models for the peak, which will
be considered later, are consistent with these observations

The temperature dependencies of the resistances of the 20.91Å and
21.12Å thick films at temperatures below 1K can be fit by an Arrhenius
form, $R=R_{0}exp(T_{0}/T)$, in fields ranging from 0 to 10 T. This
is shown in Figs. 3(a) and 3(b). The field dependencies of the activation
energy $T_{0}(B)$, and the prefactor $R_{0}(B)$, are plotted in
the lower halves of Figs. 3(c) and 3(d). The activation energy exhibits
a peak at a magnetic field close to $B_{0}$ described in the previous
paragraph. 

The measurement of resistance at temperatures below 300mK is difficult
for several reasons. The I-V characteristics become non-linear at
currents larger than 1pA. The resistance itself becomes so large that
combined with the capacitances in the measuring circuit, with its
heavy filtering, results in an extraordinarily long time constant.
Also, $R(T)$ can exceed the input impedance of the voltage amplifier,
which can lead to erroneous results. Therefore, data below 300mK were
questionable and were excluded.

\begin{figure}
\includegraphics{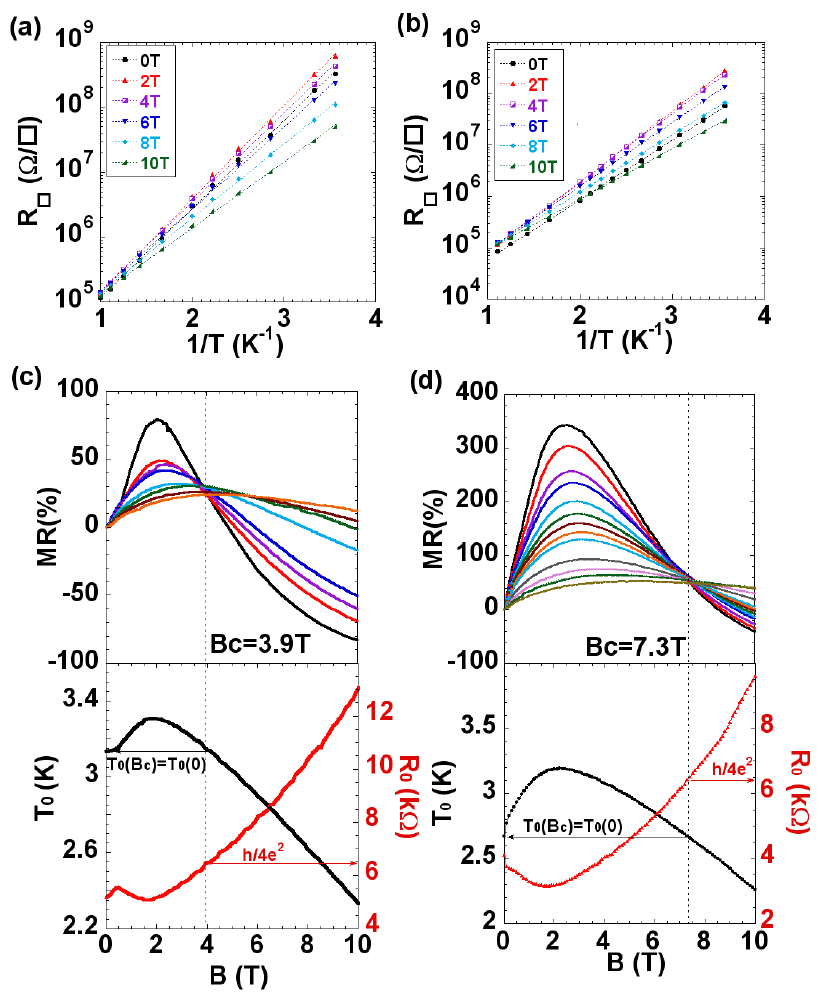}

\caption{Arrhenius plots of the (a) 20.91Å and (b) 21.12Å thick films in six
representative magnetic fields. The resistances increase by more than
three decades in these two films within the temperature range from
1K to 0.28K. The $MR$, the activation energy $T_{0},$and the prefactor
$R_{0}$ vs. magnetic field of the 20.91Å and 21.12Å thick films are
plotted in (c) and (d). The temperatures in (c) are 300mK, 400mK,
450mK, 500mK, 700mK, 800mK, 900mK, and 1K. The temperatures in (d)
are 300mK to 500mK with 25mK as the common increment and 500mK to
900mK with 100mK as the increment.}

\end{figure}

The most striking result is the occurrence of a crossover in the plot
of the $MR$ vs. $B$, as shown in Figs. 3(c) and 3(d). The magnetic
field at the crossing point, $B_{c}$, corresponds to two features
of the Arrhenius fit. First, the activation energy at this crossing
field returns to the value it exhibited at zero field. Therefore,
$T_{0}(B)-T_{0}(0)$ is always positive when $B<B_{c}$ and negative
when $B>B_{c}$. Second, the prefactors, $R_{0}$, in these two films,
are equal in value to $h/4e^{2}$, which is the quantum resistance
for electron pairs. Parenthetically the first appearance of positive
magnetoresistance at 600mK for the film thicker than 20.53Å also coincides
with the zero field prefactor falling below $h/4e^{2}$. These three
features lead us to suggest the existence of a quantum critical point
at $B=B_{c}$ with the $MR$ rather than $R$ as the observable. Indeed,
if Arrhenius conduction were to extend to zero temperature, in zero
temperature limit, we would expect

\[
MR(B,T)|_{T\rightarrow0}=\left[\frac{R_{0}(B)}{R_{0}(0)}exp(\frac{T_{0}(B)-T_{0}(0)}{T})-1\right]_{T\rightarrow0}\]

\begin{equation}
=\begin{cases}
\infty, & B<B_{c}\\
\frac{h/4e^{2}}{R_{0}(0)}, & B=B_{c}\\
-1, & B>B_{c}\end{cases}\end{equation}
even though all resistances would diverge.

Further support for the idea of a quantum phase transition comes from
the success of finite size scaling. Here we use the scaling form first
introduced by Fisher\cite{Fisher}:

\begin{equation}
R=R_{c}\mathcal{F}(\frac{|B-B_{c}|}{T^{1/\nu z}})\end{equation}
However, we use $MR$ as the observable in place of the resistance.
Both films' data, within a certain range of fields and at sufficiently
low temperatures can be scaled with critical exponent product $\nu z=0.65\pm0.08$.
This is shown in Fig. 4. With the assumption $z=1$, this product
would correspond to the universality class of a 2+1 dimensional XY
model. Similar values have been found for magnetic field and electrostatically
tuned SI transitions \cite{Markovic1,Parendo}. The data points close
to the peak in $R(B)$ and at high temperatures fail to scale, which
may due to the limits on the quantum critical regime.

\begin{figure}
\includegraphics{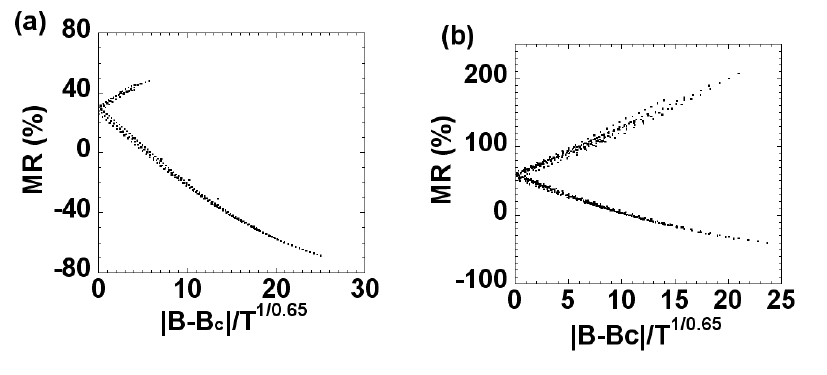}

\caption{Scaling of the $MR$ of (a) the 20.91Å and (b) the 21.12Å thick films.
Both plots employ data from below and including 500mK. The magnetic
field range of 20.91Å film is from 2.5 to 10 Tesla, while it is 5
to 10 Tesla for the 21.12Å film. In the case of the 20.91Å thick film,
there is a shorter upper branch due to the closeness of $B_{c}=3.9T$
and the peak in $R(B)$, while $B_{c}=7.3T$ in the case of the 21.12Å
thick film.}

\end{figure}

It is interesting that the $MR$ rather than the $R$ isotherms as
a function of $B$ cross as a function of magnetic field. The low
temperature zero field resistance must result from a combination of
effects including the motion of strongly localized electrons as well
as participation of presumably localized Cooper pairs. The application
of a magnetic field to the film adds vortices and the behavior of
these added vortices results in a highly resistive phase that appears
to disappear at a field-tuned quantum phase transition. That this
high resistance phase and the observed crossover are associated with
Cooper pairing is supported by the robust observation that at the
crossover magnetic field the prefactor of the Arrhenius fit to the
data is the quantum resistance for electron pairs.

Additional evidence for the presence of vortices in the film near
the magnetoresistance peak is the anisotropy of magnetoresistance.
At 400 mK for the 21.12Å thick film, our preliminary results of the
MR in a 2.5T parallel field is 20.95\%\cite{YenHsiangLin2}, while
it is 198.7\% in a 2.5T perpendicular field. This result is consistent
with previous observations by Markovic \textit{et al}.\cite{Markovic2},
which were also interpreted as the evidence of vortices in the insulating
Bi films at low magnetic fields. With the ability to apply higher
fields, we found the anisotropy diminishes when the field is larger
than the peak field and vanishes near $B_{c}$. For instance, at 400
mk the MRs in parallel and perpendicular fields of 7.3T are 56.2\%
and 60.6\% respectively. This result is consistent with the idea that
local superconductivity and vortices disappear close to to the field-tuned
quantum phase transition. 

To the best of our knowledge none of the models of the SI transition
predict a quantum phase transition such as the one reported here,
although it is quite possible that they may be extended to include
one \cite{Fisher,Ghosal,Feigelman,Galitski,Dubi,Muller,Kramer,Vinokur}.The
condition of $R_{0}$ equal to $h/4e^{2}$ delineates a phase boundary
in these thickness and field tuned insulating films as evidenced by
two observations: the magnetoresistance peak is found only in the
thicker films when the zero-field prefactor falls below $h/4e^{2}$
and the prefactor at the crossover field $B_{c}$ is $h/4e^{2}$.
This suggests that quantum fluctuations of vortices play a role in
the present observations, that $B_{c}$ is the critical field for
the vanishing of local superconductivity, and that the transition
is from a Bose insulator with localized Cooper pairs to a Fermi insulator.

One might ask why these effects have not been observed previously,
given the significant number of studies of the field-tuned superconductor-insulator
transition. Most studies have focused on films that are superconducting
in the absence of a magnetic field. Thus there is no zero-field reference
resistance as would be needed to evaluate the magnetoresistance. Secondly
the crossover field is at 3.9T and 7.3T for the two films reported
here, with the 7.3T crossover a property of the less disordered film.
With further reduction of disorder with an increase of thickness,
the crossover could move to unattainably high values of magnetic field
and be unobservable. 

In summary, isotherms of the $MR$ have been observed to cross at
a well-defined magnetic field higher than that of the peak in $R(B)$
of quench-condensed insulating films of \textit{a-}Bi. Curves of $R(T)$
at all magnetic fields follow an Arrhenius form for temperatures below
1K. The prefactor of this form becomes equal to the quantum resistance
for pairs and the activation energy returns to its zero-field value
at the crossover field. Data near the crossover are consistent with
finite size scaling and the universality class of the $(2+1)D$ XY
Model. We suggest that these observations are evidence of a quantum
phase transition between two distinct insulating phases, which might
be a Bose insulator to a Fermi insulator. 

This work was supported by the National Science Foundation under grant
NSF/DMR-0854752.

\end{document}